\documentclass[twocolumn]{aastex63}

\hypersetup{linkcolor=blue,citecolor=blue,filecolor=blue,urlcolor=blue}

\usepackage{multirow}
\newcommand{\teff}{T_{\mathrm{eff}}}
\newcommand{\lgg}{\log{g}}
\newcommand{\lggu}{\log{g / \mathrm{cm\,s^{-2}}}}
\newcommand{\feh}{\mathrm{[Fe/H]}}
\newcommand{\lgeps}[1]{\log\epsilon_{\mathrm{#1}}}
\newcommand{\xfe}[1]{\mathrm{[#1/Fe]}}
\newcommand{\vmic}{v_{\mathrm{mic}}}
\newcommand{\liline}{\ion{Li}{1} $670.8\,\mathrm{nm}$}
\newcommand{\naline}{\ion{Na}{1} $589.0\,\mathrm{nm}$}

\received{soon}
\revised{later}
\accepted{even later}
\submitjournal{ApJ}

\shorttitle{Li$^+$+H$^-$/D$^-$ and Na$^+$+H$^-$/D$^-$ collisions}
\shortauthors{Barklem et al.}

\begin{document}

\title{Mutual neutralisation in Li$^+$+H$^-$/D$^-$ and Na$^+$+H$^-$/D$^-$ collisions: Implications of experimental results for non-LTE modelling of stellar spectra}

\correspondingauthor{Paul S. Barklem}
\email{paul.barklem@physics.uu.se}
\author[0000-0003-2415-9836]{Paul S. Barklem}
\author[0000-0002-3181-3413]{Anish M. Amarsi}
\author[0000-0002-6224-3492]{Jon Grumer}
\affiliation{Theoretical Astrophysics, Department of Physics and Astronomy, Uppsala University, Box 516, SE-751 20 Uppsala, Sweden}

\author{Gustav Eklund}
\author{Stefan Ros\'en}
\author[0000-0001-8184-4595]{MingChao Ji}
\author{Henrik Cederquist}
\author[0000-0002-2493-4161]{Henning Zettergren}
\author[0000-0002-8209-5095]{Henning T. Schmidt}
\affiliation{Department of Physics, Stockholm University, Stockholm, 10691, Sweden}



\begin{abstract}

Advances in merged-beams instruments have allowed experimental studies of the mutual neutralisation (MN) processes in collisions of both Li$^+$ and Na$^+$ ions with D$^-$ at energies below 1~eV.  These experimental results place constraints on theoretical predictions of MN processes of Li$^+$ and Na$^+$ with H$^-$, important for non-LTE modelling of Li and Na spectra in late-type stars.  We compare experimental results with calculations for methods typically used to calculate MN processes, namely the full quantum (FQ) approach, and asymptotic model approaches based on the linear combination of atomic orbitals (LCAO) and semi-empirical (SE) methods for deriving couplings.  It is found that FQ calculations compare best overall with the experiments, followed by the LCAO, and the SE approaches.  The experimental results together with the theoretical calculations, allow us to investigate the effects on modelled spectra and derived abundances and their uncertainties arising from uncertainties in the MN rates.  Numerical experiments in a large grid of 1D model atmospheres, and a smaller set of 3D models, indicate that neglect of MN can lead to abundance errors of up to 0.1 dex (26\%) for Li at low metallicity, and 0.2 dex (58\%) for Na at high metallicity, while the uncertainties in the relevant MN rates as constrained by experiments correspond to uncertainties in abundances of much less than 0.01~dex (2\%).  This agreement for simple atoms gives confidence in the FQ, LCAO and SE model approaches to be able to predict MN with the accuracy required for non-LTE modelling in stellar atmospheres.


\end{abstract}


%


\section{Introduction} \label{sec:intro}

\subsection{Non-LTE modelling of stellar spectra}

The uncertainty regarding the effects of low-energy hydrogen atom collisions on
non-local thermodynamical equilibrium (non-LTE) modelling of the spectra of
late-type stars like the Sun
has been a long-standing problem \citep[e.g.][]{plaskett_interpretation_1955, Gehren1975, steenbock_statistical_1984, Lambert1993, Asplund2005, Barklem2012a, barklem_accurate_2016}.  Many initial studies both of the astrophysical modelling, and of the collision physics, focussed on Li and Na, due to their relative simplicity from an atomic physics point of view as well as their astrophysical importance.  However, the focus of initial studies was on excitation processes.  For this reason, \cite{Fleck1991} performed an experimental study of the process
\begin{equation}
\mathrm{Na}(3s) + \mathrm{H} \rightarrow \mathrm{Na}(3p, 4s) + \mathrm{H} 
\end{equation}
at moderately low collision energy (down to $\sim$10 eV) and made an analysis in terms of the avoided curve crossings in NaH potentials and the Landau-Zener model for dynamics at the crossings, showing that the data could be reasonably explained through such a model. \cite{Belyaev1999} later performed full quantum calculations (FQ, quantum chemistry with quantum dynamics) down to the threshold (2.1~eV), finding excellent agreement with the available experimental results above $\sim$10 eV.  It should be noted here that the relevant collision energies in late-type stellar atmospheres are of order 0.2 -- 0.7 eV.  

Later, \cite{Belyaev2003} made detailed quantum scattering calculations for the excitation/deexcitation processes
\begin{equation}
\mathrm{Li}(nl) + \mathrm{H} \rightleftarrows \mathrm{Li}(n^\prime l^\prime) + \mathrm{H} 
\end{equation}
including the 9 lowest states of Li up to $4f$.  They also included the charge transfer processes 
\begin{equation}
\mathrm{Li}(nl) + \mathrm{H} \rightleftarrows \mathrm{Li}^+ + \mathrm{H}^- 
\label{eq:Li_ct}
\end{equation}
based on the results of \cite{Croft1999a} and \cite{Croft1999}.  This process naturally arises from the avoided curve crossing mechanism and was found to show large cross sections compared to the excitation and deexcitation processes.  The forward reaction in equation~\ref{eq:Li_ct} is referred to as ion-pair production, and the reverse process mutual neutralisation (MN).  A complete set of rate coefficients for excitation/dexcitation and ion-pair production/mutual neutralisation processes involving Li were produced by \cite{Barklem2003b}, and applied to non-LTE modelling of stellar spectra.  This work showed that excitation/deexcitation processes were in fact unimportant in the astrophysical modelling of the Li resonance line, but the charge transfer processes had significant effects, a result later confirmed by a larger study across many different stellar atmosphere models \citep{Lind2009b}.  MN processes were later also seen to be important for Na and its spectrum \citep{Belyaev2010, Barklem2010, Lind2011}.

At the time of these studies, it was difficult to estimate the uncertainties in the MN cross sections in the energy regime relevant to late-type stellar atmospheres, roughly 0.2 -- 0.7 eV, and thus in the rate coefficients for relevant temperatures, roughly 2000 -- 7000~K. \cite{PeartMeasurementsmutualneutralisation1987} had measured total cross sections for MN in Li$^+$+H$^-$ collisions in an inclined-beams experiment carried out in Newcastle, but at energies above 33 eV.  \cite{peart_merged_1994} later measured MN in Li$^+$+D$^-$ collisions down to $\sim 0.7$~eV in a merged-beams experiment.  \cite{Croft1999a} performed calculations for MN in both Li$^+$+H$^-$ and Li$^+$+D$^-$, which allowed direct comparison with the experimental results for Li$^+$+D$^-$ of \cite{peart_merged_1994} at low energy.  After a correction for the finite aperture of the detectors in the experiment, the results were found in reasonable agreement, differing by about 10\% at energies in the range 10 -- 50 eV, and the deviation increased with decreasing energy to reach 20\% at the lowest measured energy of around 0.7 eV.  However, this comparison does not probe the energy range of most interest in stellar atmospheres, a few 0.1 eV. \cite{Croft1999} extended their calculations for Li$^+$+H$^-$ to lower energies (meV energies), allowing calculations of rate coefficients for temperatures of a few thousand degrees.  

Based on the poorer agreement with experiment for Li$^+$+D$^-$ at lower collision energies, \cite{Barklem2003b} performed numerical experiments for three stars assuming an error of 50\%.  For example, in the classical metal-poor star HD140283 it was found that including the calculated MN processes, compared to ignoring them, changed the \ion{Li}{1} resonance line strength by 20\%, while if half of the MN rates are used the line strength changed by 18\%.  Thus, the MN rates do not influence the modelled line strength linearly (due to the general non-linear behaviour of the statistical equilibrium equations and competition with other processes), and the latter was found to be relatively insensitive to the uncertainty in the MN rate.  However, this uncertainty would still give uncertainties in line strengths and abundances of a few per cent, and was not well constrained.  Given that spectral lines are regularly measured to better than per cent level, and that more accurate abundances will yield more information, it is useful to constrain this uncertainty further.  Further, this numerical experiment only probes the effect of uncertainty in the total rate, but not the possible effect of changes in the branching fractions between final states.

\subsection{Measurements of mutual neutralisation and comparisons with atomic theory}

Recent advances in merged-beams instruments have opened the way for experimental studies of MN at low energies ($<1$~eV).  In particular, for the process
\begin{equation}
\mathrm{Li}^+ + \mathrm{D}^- \rightarrow \mathrm{Li}(3s,3p,3d) + \mathrm{D},
\label{eq:Li_mn}
\end{equation}
total cross sections and branching fractions have been measured in Louvain \citep{LaunoyMutualNeutralizationLi2019}, and branching fractions in Stockholm \citep{EklundCryogenicmergedionbeamexperiments2020} at the DESIREE infrastructure \citep{thomas_double_2011, SchmidtFirststorageion2013}.  Results for branching fractions have also been obtained for 
\begin{equation}
\mathrm{Na}^+ + \mathrm{D}^- \rightarrow \mathrm{Na}(4s,3d,4p) + \mathrm{D}
\label{eq:Na_mn}
\end{equation} reactions at DESIREE (Eklund et al.~in prep.).  We note here that the low energy experimental results using merged-beams instruments from Louvain, DESIREE, as well as the earlier results from Newcastle, all consider MN predominantly with D$^-$, despite that it is H$^-$ that is of interest for astrophysics.  The maximum mass ratio for any given merged-beams instrument is determined by the ratio of the highest to the lowest energy ion beams that can be handled.  For the existing instruments the value of this ratio is of the order of 10 for low (near zero eV) relative collision energies. For the specific case of DESIREE, the limit is a factor of 20 as reported in the design paper \citep{thomas_double_2011}.  Though modern merged-beams experiments can in principle handle Li$^+$+H$^-$ (in fact such an experiment was also performed for this case in Louvain for one energy as a check on the Li$^+$+D$^-$ branching fraction results) there are various advantages to using D$^-$, including that if the masses are more similar it is easier to produce beams of the same velocity (leading to low relative velocity and thus low energy collisions).  In addition, heavier isotopes are less affected by stray magnetic fields. 

The new experimental results at low energies allow us to constrain the uncertainties in the rate coefficients used in non-LTE modelling of stellar spectra with much greater precision than before.  Further, since the theoretical and astrophysical modelling studies mentioned above in the period 2003--2011, two asymptotic model approaches to the problem of hydrogen collisions have been developed, both based on Landau-Zener dynamics of the avoided ionic curve crossing mechanism, but with different methods for calculating the couplings at the avoided crossings.  The first method, developed by \cite{Belyaev2013}, uses semi-empirical (SE) couplings from \cite{Olson1971}.   The second method, developed by \cite{barklem_excitation_2016, barklem_erratum:_2017}, employs a two-electron linear combination of atomic orbitals (LCAO) method, extending the method of  \cite{Grice1974} and \cite{Adelman1977}.  These asymptotic model approaches have allowed estimates of hydrogen collision rates, including MN, for many astrophysically interesting and complex atoms, such as Al \citep{Belyaev2013}, Ca \citep{barklem_excitation_2016}, Mn \citep{belyaev_atomic_2017, GrumerExcitationchargetransfer2020}, Fe \citep{BarklemExcitationchargetransfer2018}, C, N, \citep{AmarsiExcitationchargetransfer2019} and O \citep{BarklemExcitationchargetransfer2018a}, among others.  

In view of these theoretical developments, as well as
the new experimental results, we now approach the question of 
the uncertainty in MN rates, and the direct implications for modelling Li and
Na spectra. We will discuss how the new information may shed light on the accuracies of the asymptotic model approaches in general (Sect.~\ref{sec:exp}).  Further, 
with increased computing power, as well as
continued development of radiative transfer codes, it is now feasible to
perform non-LTE spectrum synthesis across very large grids of 1D or small
grids of 3D stellar atmospheres to assess the impact of the uncertainties in MN
rates more thoroughly (Sect.~\ref{sec:comp}).  Finally the conclusions are presented and in particular it is found that though there are some differences between the theoretical predictions from the FQ, LCAO and SE methods, and some discrepancies with the experimental results, the methods are all sufficiently accurate (based on comparison with experiment) and sufficiently similar that the choice of theoretical MN description among these methods for non-LTE modelling of Li and Na spectra in late-type stars has only small effects on the modelled spectral lines, and thus small effects on derived abundances. 


\section{Experimental results and comparison with theory} \label{sec:exp}

As discussed above, the present experimental results are for MN with D$^-$, rather than the case of actual astrophysical interest H$^-$, and this has consequences for our comparisons.  The isotopic differences between the XH and XD (here X is Li or Na) interaction potentials and couplings due to the mass (e.g. nuclear motion, mass polarisation) and nuclear spin differences (e.g spin orbit coupling) are expected to be small.  These effects are often neglected in molecular structure calculations, as was done by \cite{Croft1999a} in their calculations for LiH and LiD where they used the same potentials for Li$^+$+H$^-$ and Li$^+$+D$^-$.  Regarding the collision dynamics, at high energies cross sections are expected only to be a function of the relative collision velocity, and thus isotope effects are also negligible.  However, trajectory effects arise due to the mass difference at low energies.  
In particular, Coulomb focussing causes the cross sections for processes with H$^-$ to be larger than those for D$^-$, since for given values of the relative energy and impact parameter the lighter anion will have a shorter distance of closest approach to the cation.  Since partial cross sections into different final states are affected differently, this affects the branching fractions.  Note that quantum dynamical calculations will also show interference effects that may differ between masses due to the trajectory effects.

As it is H$^-$ that is of interest in astrophysics, the FQ calculations that have been performed so far have often focussed on MN with H$^-$.  In the case of Li, while \cite{Croft1999a} calculated for Li$^+$+D$^-$ down to 0.68 eV to compare with the experimental results of \cite{peart_merged_1994}, \cite{Croft1999} calculated for Li$^+$+H$^-$ down to meV energies to enable rate coefficients to be calculated.  For Na$^+$+H$^-$, \cite{dickinson_initio_1999} have made FQ calculations down to meV energies, while no FQ calculations are available for Na$^+$+D$^-$.   For this reason, it is useful to be able to present experimental and theoretical results for MN with D$^-$ and H$^-$ together, and to present the comparisons on the reduced energy scale, rather than with respect to absolute collision energy.  The reduced energy is defined as $E_\mathrm{R} = E_\mathrm{CM}/\mu = \frac{1}{2}v^2$, where $E_\mathrm{CM}$ is the collision energy in the centre-of-mass frame, $\mu$ is the reduced mass of the system\footnote{The reduced masses are for Li+H 0.88, Li+D 1.56, Na+H 0.97, and Na+D 1.85 amu.}, and $v$ is the relative velocity.    On this scale the collisions for both D$^-$ and H$^-$ have the same relative velocity $v$ at the same $E_\mathrm{R}$, and this scaling has an additional advantage of being independent of reference frame.  At high $E_\mathrm{R}$ the results for H$^-$ and D$^-$ should agree on this scale, while at low $E_\mathrm{R}$, trajectory effects give rise to differences.  

Comparison of calculations  for Li$^+$+H$^-$ and Li$^+$+D$^-$ with the LCAO model show the effect of Coulomb focussing on the total cross section is 4, 25, and 54 per cent at $E_\mathrm{R}=$10, 1, and 0.1 eV, respectively (see below and Fig.~\ref{fig:Li_total}), while the effects on branching fractions at low energy are typically 5 per cent (in absolute terms) or less (see below and Fig.~\ref{fig:Li_bf}).  For Na$^+$+H$^-$ and Na$^+$+D$^-$ the effects are similar: 4, 25, and 61 per cent at $E_\mathrm{R}=$10, 1, and 0.1 eV, respectively (see below and Fig.~\ref{fig:Na_total}), and for branching fractions are also typically 5 per cent or less (see below and Fig.~\ref{fig:Na_bf}).

Our comparisons will focus on available FQ results, as well as the LCAO and SE asymptotic models, as these are the approaches that have been used to generate rate coefficients and applied in astrophysics, and for which total and partial FQ cross sections as a function of collision energy are available or can be calculated by us with existing codes based on LCAO or SE methods.  Results from the SE model presented here are calculated by us with the same code as used for the LCAO model, such that the only difference between the two sets of calculations is that the couplings are calculated using the semi-empirical formula of \cite{Olson1971} in the SE case.  The SE model approach used by Belyaev and collaborators uses this formula for the couplings, but naturally their descriptions and codes will differ slightly in other ways.  In any case, to our knowledge they have not made any calculations for Li and Na, and further this approach allows us to separate out the key effects of different ways to calculate the coupling strength.

\subsection{Li$^+$ + H$^-$/D$^-$}

\cite{LaunoyMutualNeutralizationLi2019} have performed merged-beams
experiments for Li$^+$+D$^-$ in Louvain, obtaining total MN cross sections for a
large number of collision energies, $E_\mathrm{CM}$, between 3.9~meV and
1.1~eV, as well as branching fractions into the $3s$, $3p$ and $3d$ states at
three energies, $E_\mathrm{CM} = $ 3.9, 20, and 200~meV.  They also obtained
branching fractions for Li$^+$+H$^-$ at $E_\mathrm{CM} = 3$~meV, which are in agreement with their Li$^+$+D$^-$ values, though no error estimates are given. 
\cite{EklundCryogenicmergedionbeamexperiments2020} have with DESIREE obtained branching
fractions into the $3s$ state for Li$^+$+D$^-$ MN at $E_\mathrm{CM} = $ 78, 262,
and 630~meV.  The $3p$ and $3d$ states are not resolved, but the measured $3s$ branching fraction (57.8$\pm$0.7 \%) at 78 meV is the most precisely measured branching fraction value.  The combined experimental results cover the energy range of interest for late-type stellar atmospheres (200-700 meV) rather well.  

Before comparing with the FQ, LCAO and SE calculations, we note that \cite{LaunoyMutualNeutralizationLi2019} made extensive calculations using four sets of quantum chemical calculations, combined with Landau-Zener dynamics, and compared with their experimental results.  They showed that the branching fractions obtained are sensitive to the quantum chemistry calculations (see their Fig.~6).  They obtained very good agreement with the branching fractions and total cross sections at low energy with their calculation using ACV5Z+G quantum chemical data.

\begin{figure}[t!]
\center
\includegraphics[width=0.49\textwidth,angle=0,trim={0 25mm 0 0},clip]{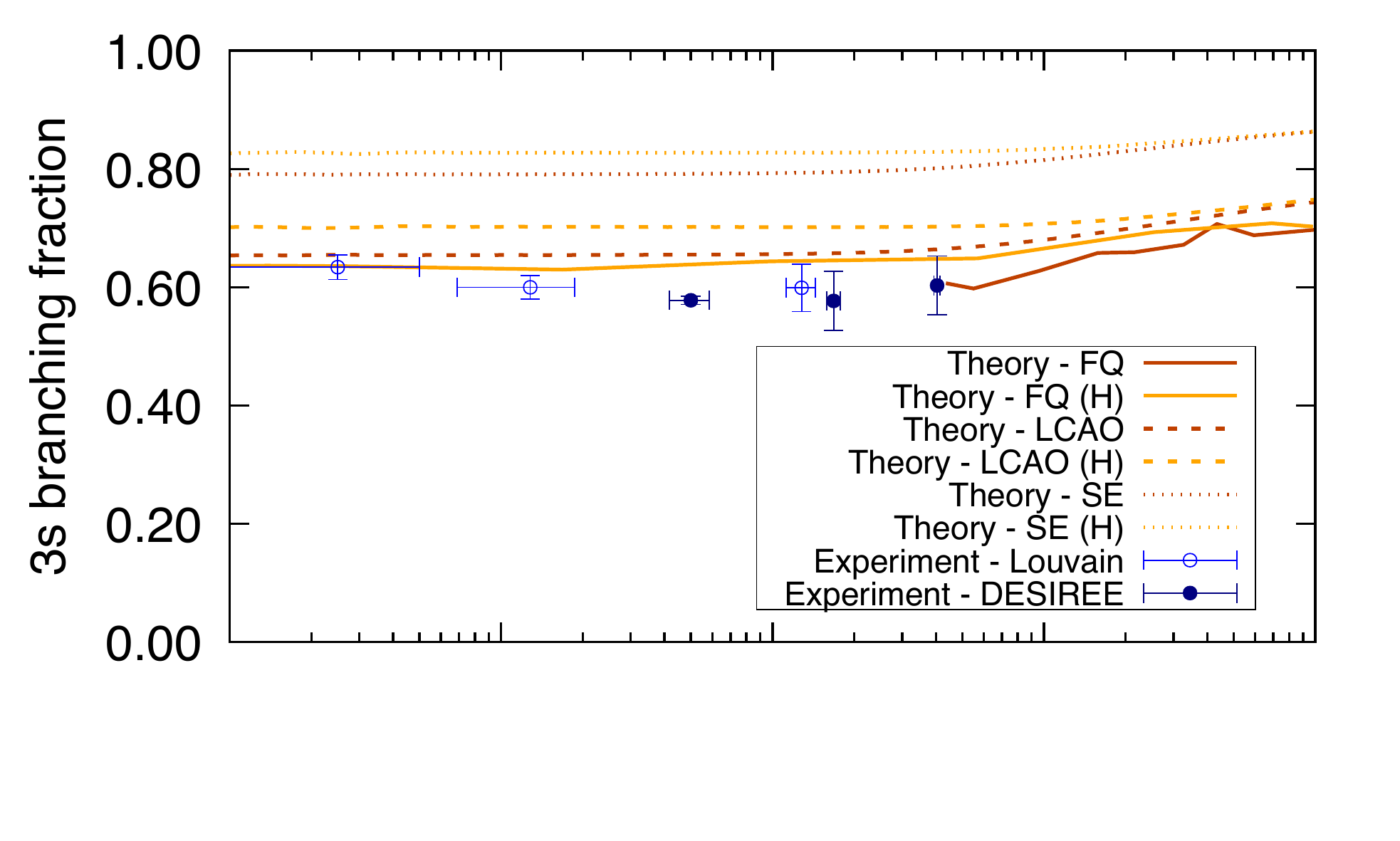}
\includegraphics[width=0.49\textwidth,angle=0,trim={0 25mm 0 0},clip]{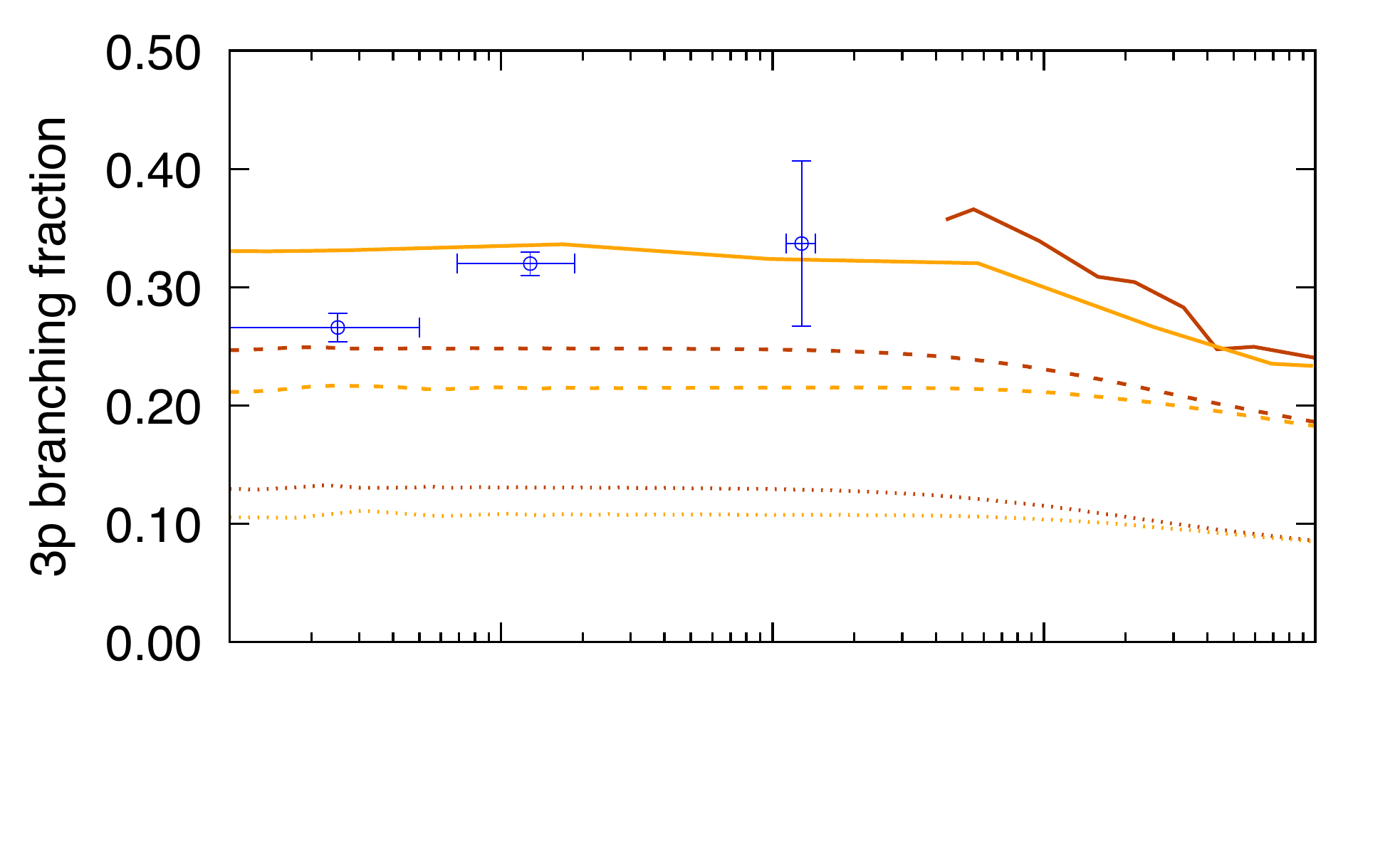}
\includegraphics[width=0.49\textwidth,angle=0]{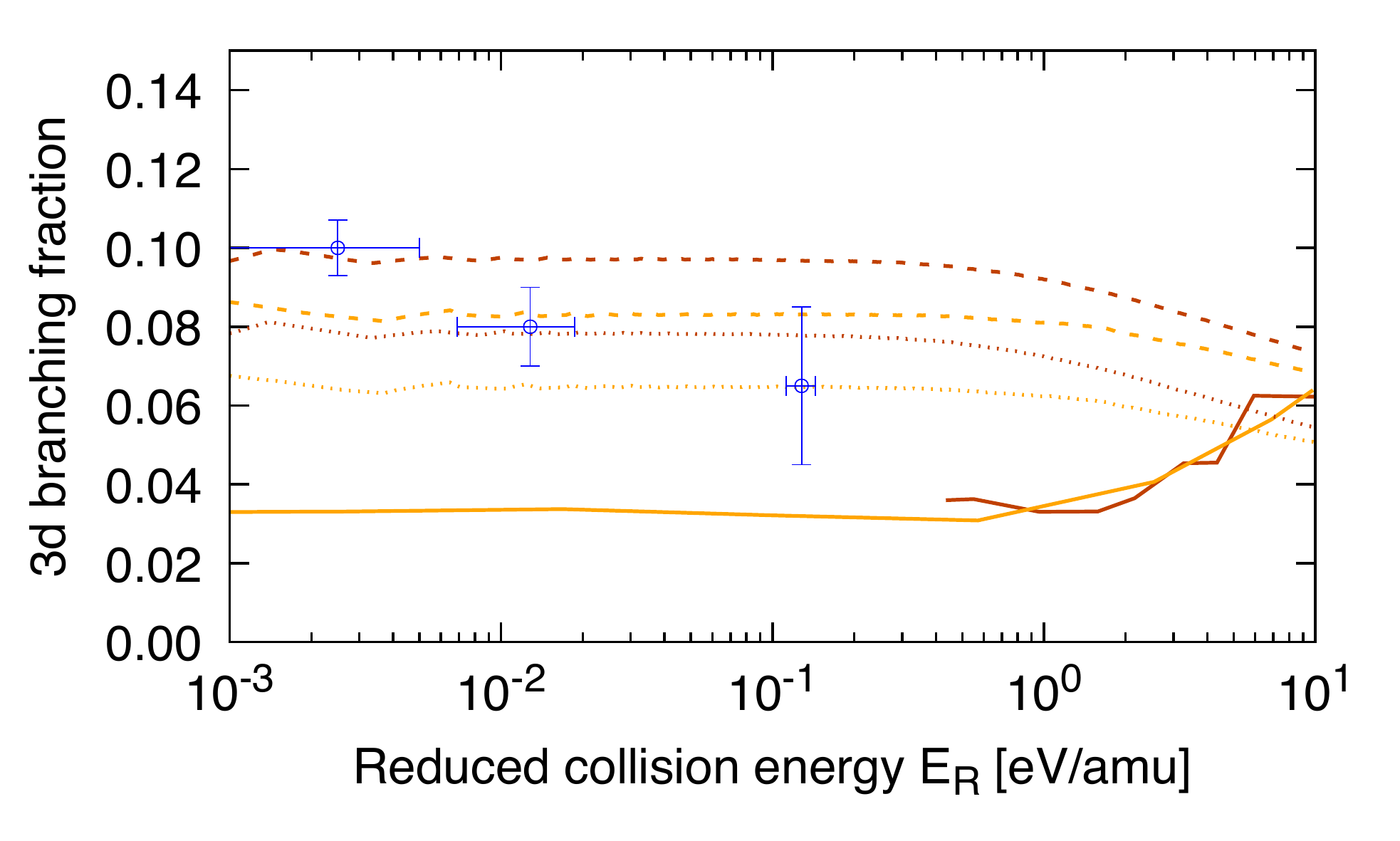}
\caption{Branching fractions for the MN reaction $\mathrm{Li}^+ + \mathrm{D}^- \rightarrow \mathrm{Li}(nl) + \mathrm{D}$ as a function of reduced collision energy.  The $3s$, $3p$ and $3d$ channels are shown in separate panels.  Experimental results from Louvain \citep{LaunoyMutualNeutralizationLi2019} and from DESIREE \citep{EklundCryogenicmergedionbeamexperiments2020} are shown, with estimated errors ($1\sigma$).  Theoretical results are shown from FQ calculations \citep{Croft1999a}, and from our calculations using both the LCAO method \citep{barklem_excitation_2016} and semi-empirical (SE) couplings from \cite{Olson1971}.  Theoretical results are also shown for the case where D$^-$ is replaced by H$^-$, marked by (H) in the legend. For Li$^+$+H$^-$ the FQ results are from \cite{Croft1999}. \label{fig:Li_bf}}
\end{figure}

In Fig.~\ref{fig:Li_bf} the experimental branching fractions are compared with
FQ, LCAO and SE theoretical calculations.  The branching fractions from Louvain and DESIREE for
the $3s$ production are in reasonable agreement, and together imply a rather flat
trend at low energy, in good agreement with the theoretical
predictions.  The FQ calculation of \cite{Croft1999a} for
Li$^+$+D$^-$ agrees very well with the DESIREE result at $E_\mathrm{CM} = 623$
meV ($E_\mathrm{R} = 404$ meV/amu), but unfortunately the calculations do not
extend to lower energies.  The Li$^+$+H$^-$ branching fraction for $3s$ from
\cite{Croft1999} is generally of order 0.05 higher than the calculation for Li$^+$+D$^-$, the result of Coulomb focussing.   

\begin{figure}[t!]
\includegraphics[width=0.49\textwidth,angle=0]{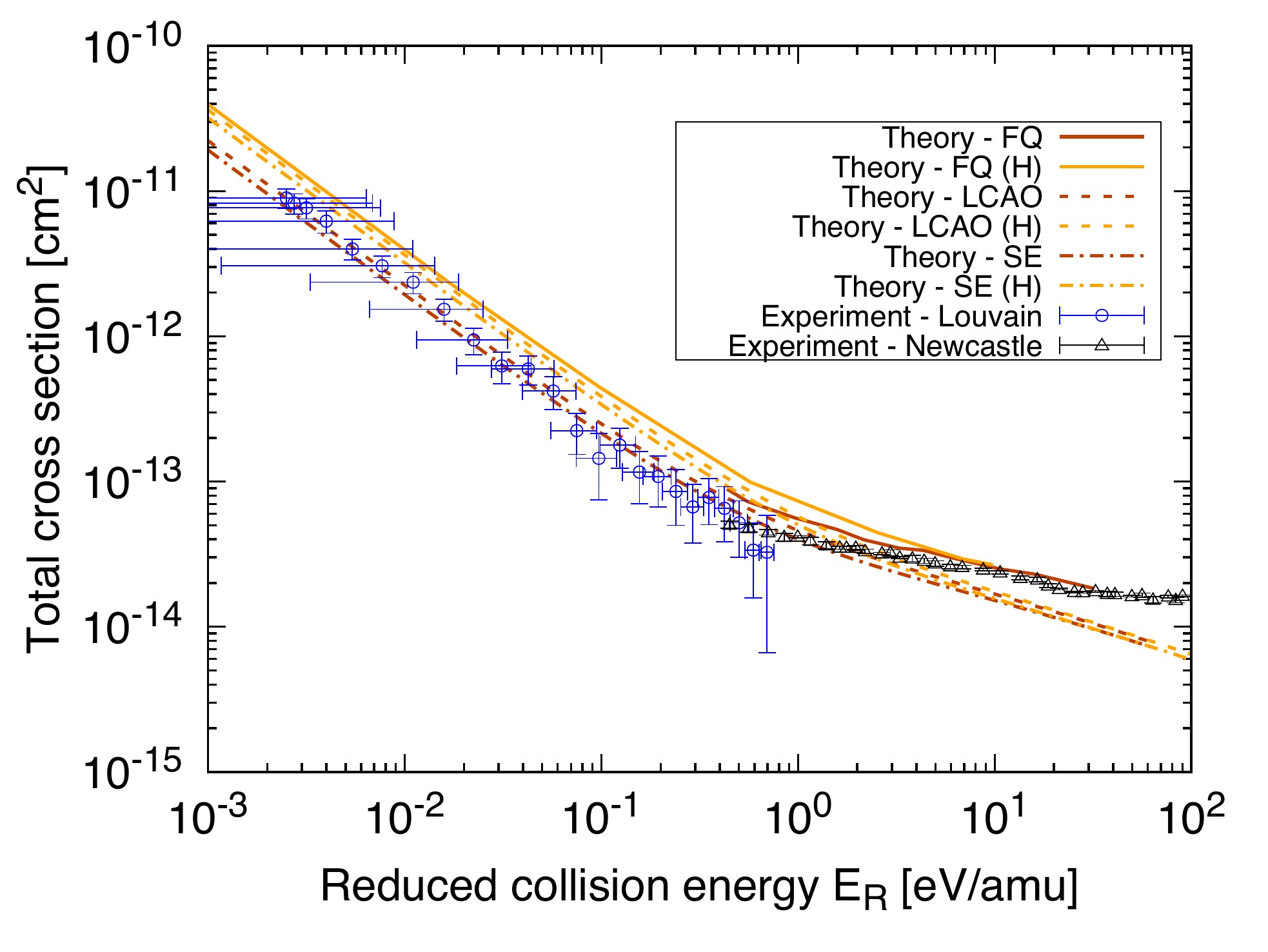}
\caption{Total cross sections for the MN reaction $\mathrm{Li}^+ + \mathrm{D}^- \rightarrow \mathrm{Li} + \mathrm{D}$ as a function of reduced collision energy.  Experimental results from Louvain \citep{LaunoyMutualNeutralizationLi2019} and from Newcastle \citep{peart_merged_1994} are shown, with estimated errors ($1\sigma$).  Theoretical results are shown from FQ calculations \citep{Croft1999a}, and from our calculations using both the LCAO method \citep{barklem_excitation_2016} and semi-empirical (SE) couplings from \cite{Olson1971}.  Theoretical results are also shown for the case where D$^-$ is replaced by H$^-$, marked by (H) in the legend. For Li$^+$+H$^-$ the FQ results are from \cite{Croft1999}. \label{fig:Li_total}}
\end{figure}

In Fig.~\ref{fig:Li_total}, the experimental total cross sections of \cite{LaunoyMutualNeutralizationLi2019}, and those at higher energies from \cite{peart_merged_1994}, are compared with calculations.  It should be noted that the \cite{peart_merged_1994} results plotted do not include the finite aperture correction implied by \cite{Croft1999a}, which would increase the total cross section by roughly 30\% at the lowest measured energy.  The FQ results for Li$^+$+D$^-$ of \cite{Croft1999a} are generally larger than both sets of experimental results in the region $E_\mathrm{R} = 0.5 - 1$ eV/amu, and rough extrapolation to lower energy would suggest this difference remains at $E_\mathrm{R} \lesssim 0.5$ eV/amu where the cross section scales as $E^{-1}$.  Correction of the Li$^+$+H$^-$ FQ results for Coulomb focussing at $E_\mathrm{R} \lesssim 0.5$ eV/amu in line with the magnitude implied by the LCAO and SE results, would also suggest that the FQ results are larger than the experimental results.  Together these comparisons suggest the FQ total cross section is of order 10-30\% larger than the experimental total cross sections, but still agreeing within the experimental uncertainties.  

Generally, the total cross sections from the FQ, LCAO and SE models are all close to the Louvain experimental results at low energy, though the FQ results deviate the most.  Based on the scatter between the different calculations, and the general agreement with experiment, it seems reasonable to conclude that the uncertainty in the total cross section from any given theoretical calculation is not worse than 30\%.  Perhaps even more importantly, the comparisons indicate that the uncertainty is roughly constant in the region of  $E^{-1}$ behaviour; the growing difference between theory and experiment seen in \cite{Croft1999a} was a result of the lowest measured energies being in the ``knee'' region between the $E^{-1}$ behaviour of the cross section at low energy, and the flatter behaviour at high energy.  The FQ results agree best with the $3s$ and $3p$ branching fractions, followed by the LCAO model, while the SE model results perform worst, often disagreeing by of order 0.2.  For $3d$ none of the theoretical models match the experimental results.  
Considering all the comparisons together, we suggest that the strongest experimental constraint is the $3s$ branching fraction, as there are two independent experimental studies, and the experimental uncertainties are significantly smaller than the differences between the models.  For Li$^+$+H$^-/$D$^-$ we conclude that the physically most advanced FQ theory performs best while the less sophisticated LCAO and SE models are somewhat less successful.  

\subsection{Na$^+$ + H$^-$/D$^-$}

\begin{figure}[t!]
\includegraphics[width=0.49\textwidth,angle=0,trim={0 25mm 0 0},clip]{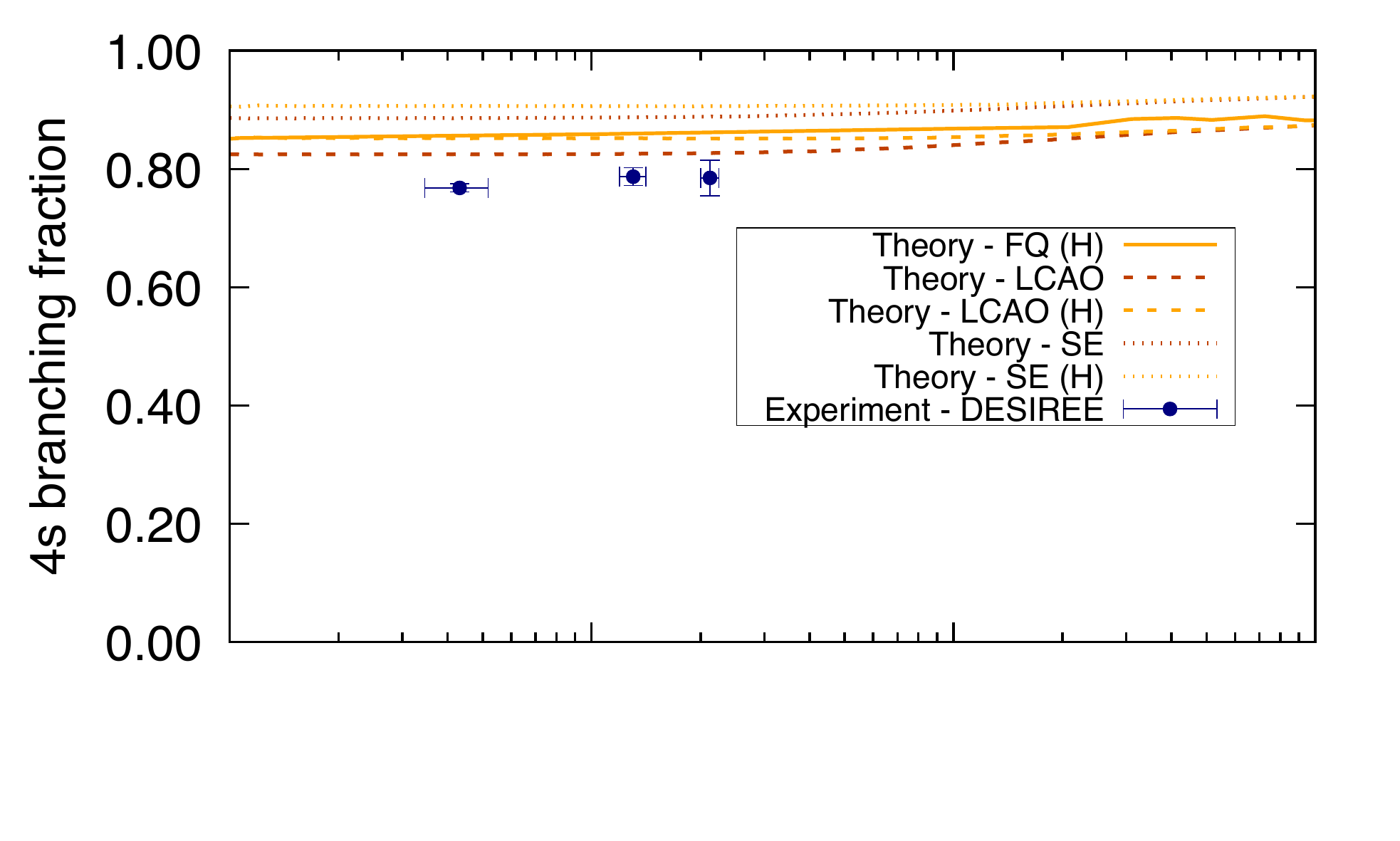}
\includegraphics[width=0.49\textwidth,angle=0,trim={0 25mm 0 0},clip]{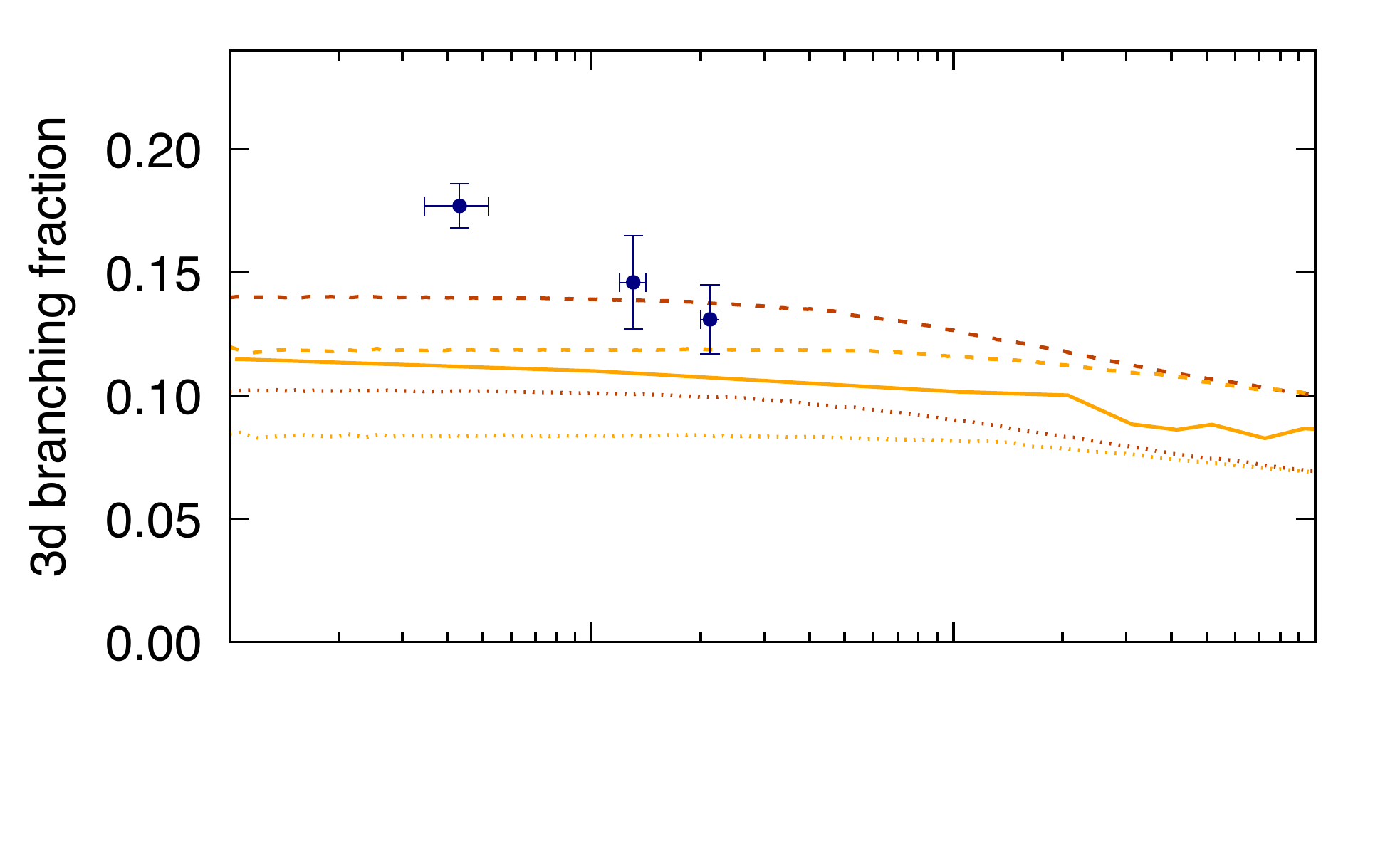}
\includegraphics[width=0.49\textwidth,angle=0]{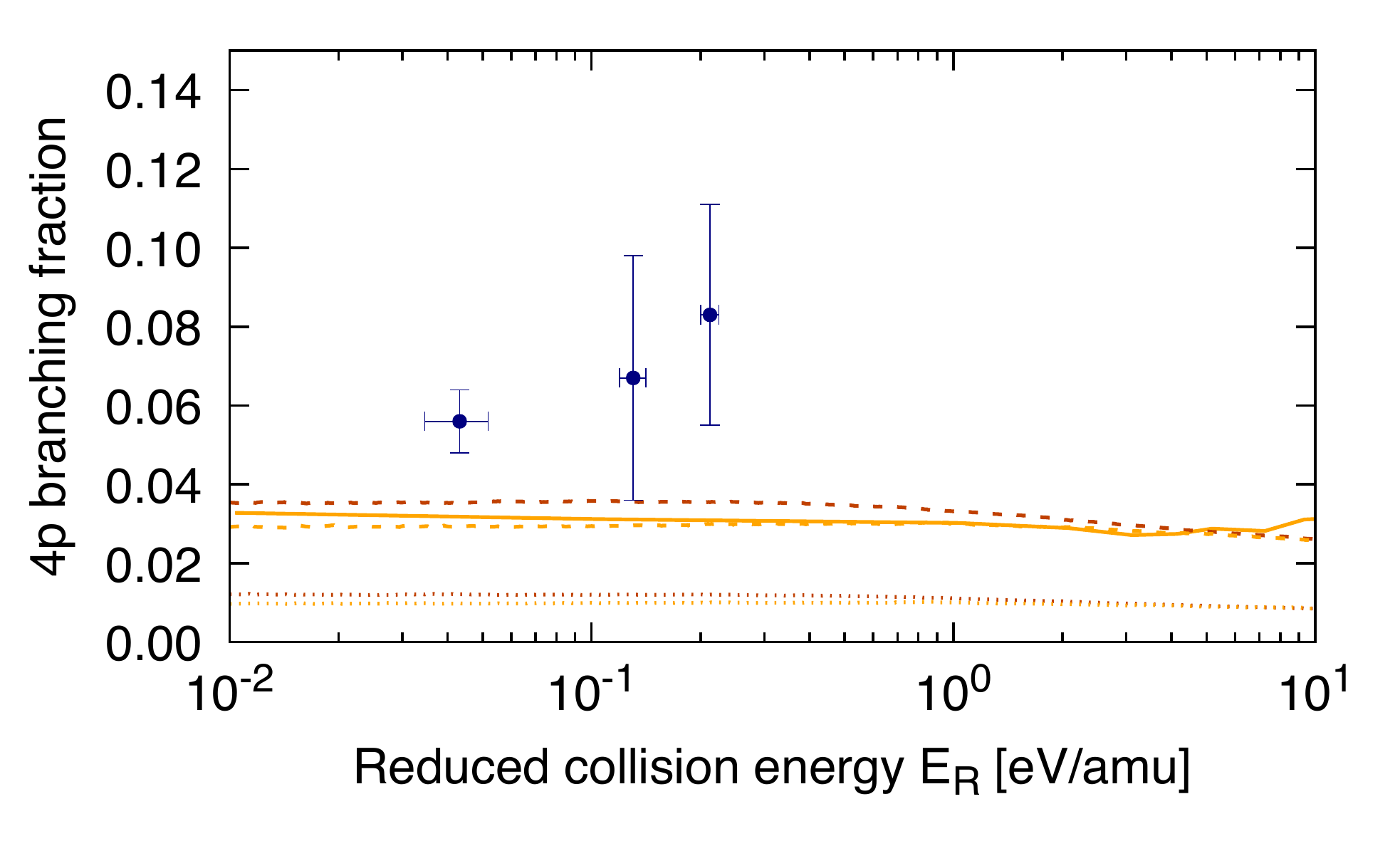}
\caption{Branching fractions for the MN reaction $\mathrm{Na}^+ + \mathrm{D}^- \rightarrow \mathrm{Na}(nl) + \mathrm{D}$ as a function of reduced collision energy.  Experimental results from DESIREE (Eklund et al. in prep.) are shown, with estimated errors ($1\sigma$).  Theoretical results are shown from our calculations using both the LCAO method \citep{barklem_excitation_2016} and semi-empirical (SE) couplings from \cite{Olson1971}.  Theoretical results are also shown for the case where D$^-$ is replaced by H$^-$, marked by (H) in the legend.  FQ results for H$^-$ are from  \cite{dickinson_initio_1999}.  \label{fig:Na_bf}}
\end{figure}

For Na$^+$+D$^-$ MN reactions, branching fractions have been obtained with DESIREE for the production of $4s$, $3d$, and $4p$ final states at three energies, $E_\mathrm{CM} = 79, 240, 392$~meV (Eklund et~al.~in prep.).  All three states are resolved, though we note that at higher energy the error bars are larger due to the decreased resolution and thus increased difficulty to resolve $3d$, and $4p$.  The obtained branching fractions are compared with FQ, LCAO and SE theoretical calculations for both Na$^+$+D$^-$ and Na$^+$+H$^-$ in Fig.~\ref{fig:Na_bf}.  For Na$^+$+D$^-$ there are no FQ calculations, while there are FQ calculations for Na$^+$+H$^-$ \citep{dickinson_initio_1999}.  On the assumption that the differences between Na$^+$+D$^-$ and Na$^+$+H$^-$ branching fractions for LCAO and SE calculations reflect the size of the Coulomb focussing effect, one expects that FQ calculations for Na$^+$+D$^-$ would lead to a lowering of the $4s$ branching fraction compared to that seen for Na$^+$+H$^-$, and an increase in the $3d$ and $4p$ branching fractions.  In the $4s$ and $4p$ cases the expected correction is not sufficient to bring the theoretical predictions into agreement with the experimental results.  In particular, for the $4s$ branching all theoretical predictions are significantly higher than the experimental results, the lowest energy value for $4s$ differing by several standard deviations from any of the theoretical values.  The experiment, however, confirms the ordering of relative populations of states predicted by all three theoretical calculations, namely MN predominantly populates the $4s$ channel, followed by $3d$, and finally $4p$.   Though there are significant discrepancies for the FQ and the LCAO results, they clearly perform better than those from the SE model for all branching fractions measured. \cite{Janev1978}, using the Landau-Herring method, obtained the surprising result that $3p$ has the second largest branching fraction among final channels, with a branching fraction of order 0.1, and this result is ruled out by the experimental results from DESIREE, where no capture into $3p$ is detected. 

\begin{figure}[t!]
\includegraphics[width=0.49\textwidth,angle=0]{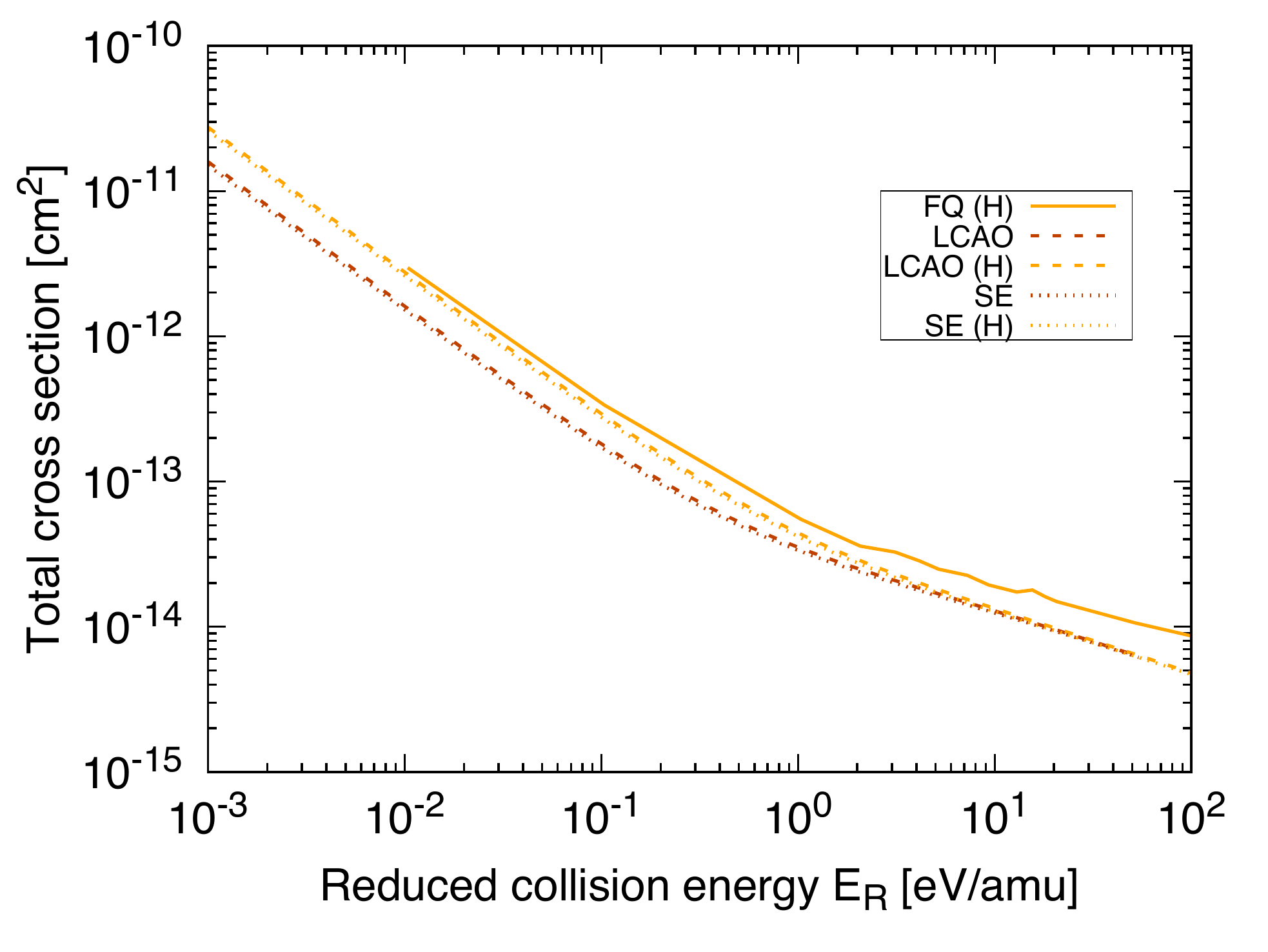}
\caption{Total cross sections for the MN reaction $\mathrm{Na}^+ + \mathrm{D}^- \rightarrow \mathrm{Na} + \mathrm{D}$ as a function of reduced collision energy. Theoretical results are shown from our calculations using both the LCAO method \citep{barklem_excitation_2016} and semi-empirical (SE) couplings from \cite{Olson1971}.  Theoretical results are also shown for the case where D$^-$ is replaced by H$^-$, marked by (H) in the legend. FQ results for H$^-$ are from  \cite{dickinson_initio_1999}.  \label{fig:Na_total}}
\end{figure}

While there are no experimental results for the total cross section for MN involving Na, the theoretical results are shown in Fig.~\ref{fig:Na_total}.  Here one sees that the general relationship between the different models is similar to the case of Li.  In particular, the effect on the total cross section due to Coulomb focussing predicted by the LCAO and SE models is roughly the same, and the FQ calculations provide larger total cross sections than the LCAO and SE models.  

In general, the comparisons for branching fractions in Na are in line with the overall results for Li, with the FQ results (correcting for Coulomb focussing) and the LCAO results generally performing better than those from the SE model. Considering the Li and Na results together, but excluding the Li $3d$ branching fraction results which are not well explained by any of the theoretical calculations, the FQ results match experiments the best, followed by the LCAO model, while the SE model performs least well of the three theoretical approaches, sometimes giving branching fractions discrepant by of order 0.05-0.2.  Finally, we note that though the experimental results for branching fractions in Li($3p$), Li($3d$), Na($3d$), and Na($4p$) show some hint of a possible trend with energy, noting that the error bars are $1\sigma$, the experimental results are also compatible with a flat trend as predicted by theory.  More, or more precise, results would be required to study this question further.

\section{Implications for non-LTE modelling of stellar spectra} \label{sec:comp}

\begin{deluxetable*}{llll}
\tablecaption{Model atoms for Li and Na. 
    Describes the rate coefficient data used 
    for MN involving the \{$3s$, $3p$, $3d$\} states of Li, and 
    for MN involving the \{$4s$, $3d$, $4p$\} states of Na.
    \label{tab:modelatoms}}
\tablewidth{0pt}
\tablehead{
    \colhead{Element} &
    \colhead{ID} & \colhead{Description} & 
    \colhead{Reference} 
}
\startdata
    \multirow{6}{*}{Li/Na} & FQ & Quantum chemistry potentials + Full quantum scattering  & 
    1, 2, 3, 4, 5, 6 \\
    & LCAO &  Theoretical LCAO couplings + Landau-Zener model & 7 \\
    & SE & Semi-empirical couplings + Landau-Zener model & 8 \\ 
    & SE-sFQ & SE scaled to have FQ total rate & - \\
    & None & Zero rates & - \\
    \hline
    \multirow{2}{*}{Li} & 
    BV18 & Quantum chemistry potentials + Quantum current probability method & 
        9, 10 \\
    & L19 & Quantum chemistry potentials + Landau-Zener model & 11 \\ 
\enddata
    \tablerefs{(1)~\cite{Croft1999};
    (2)~\cite{Belyaev2003};
    (3)~\cite{Barklem2003b};
    (4)~\cite{dickinson_initio_1999};
    (5)~\cite{Belyaev2010};
    (6)~\cite{Barklem2010};
    (7)~\cite{barklem_excitation_2016};
    (8)~\cite{Olson1971};
    (9)~\cite{Croft1999a};
    (10)~\cite{BelyaevAtomicDataInelastic2018};
    (11)~\cite{LaunoyMutualNeutralizationLi2019}.}
\end{deluxetable*}

\begin{figure}[t]
    \includegraphics[scale=0.33]{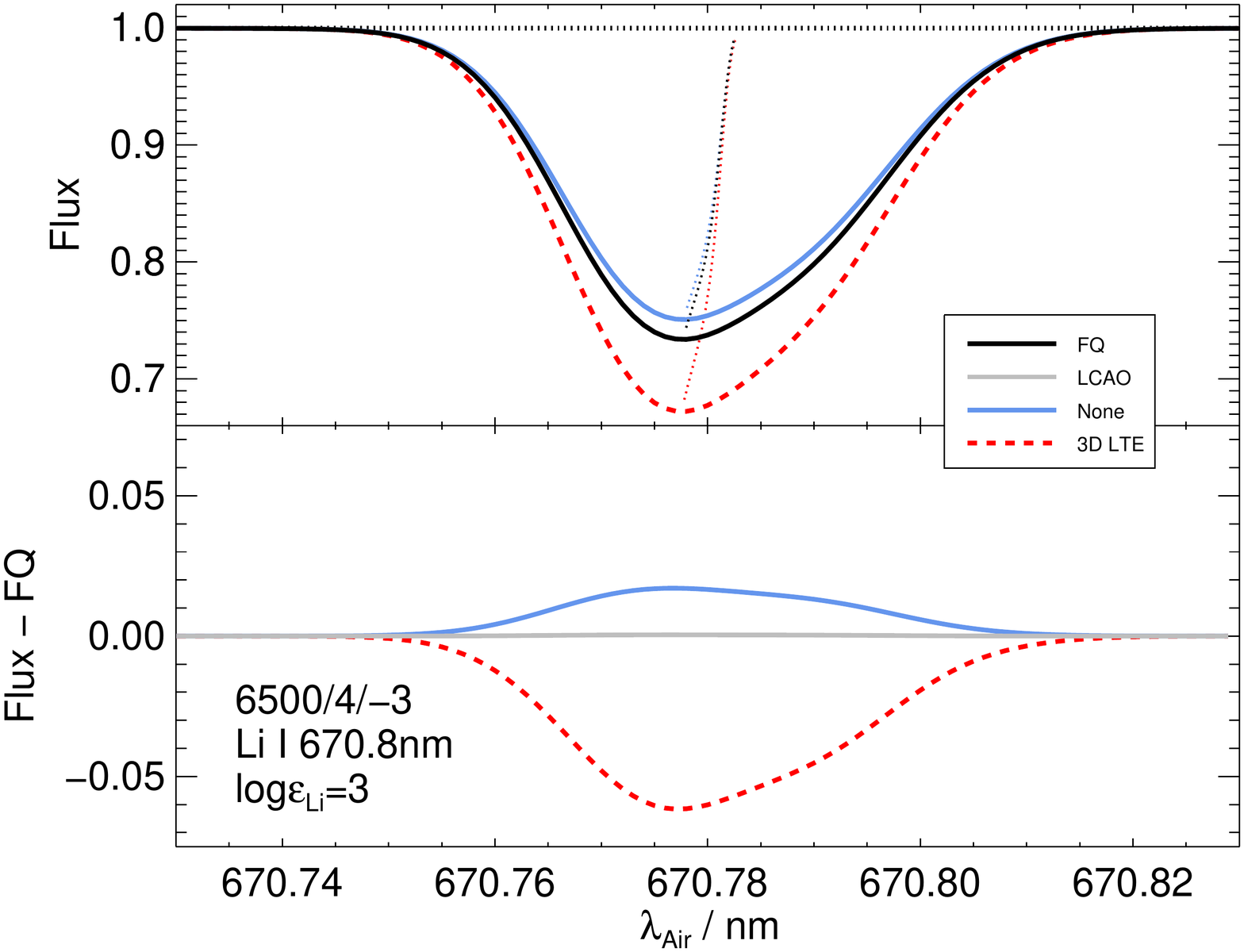}
    \includegraphics[scale=0.33]{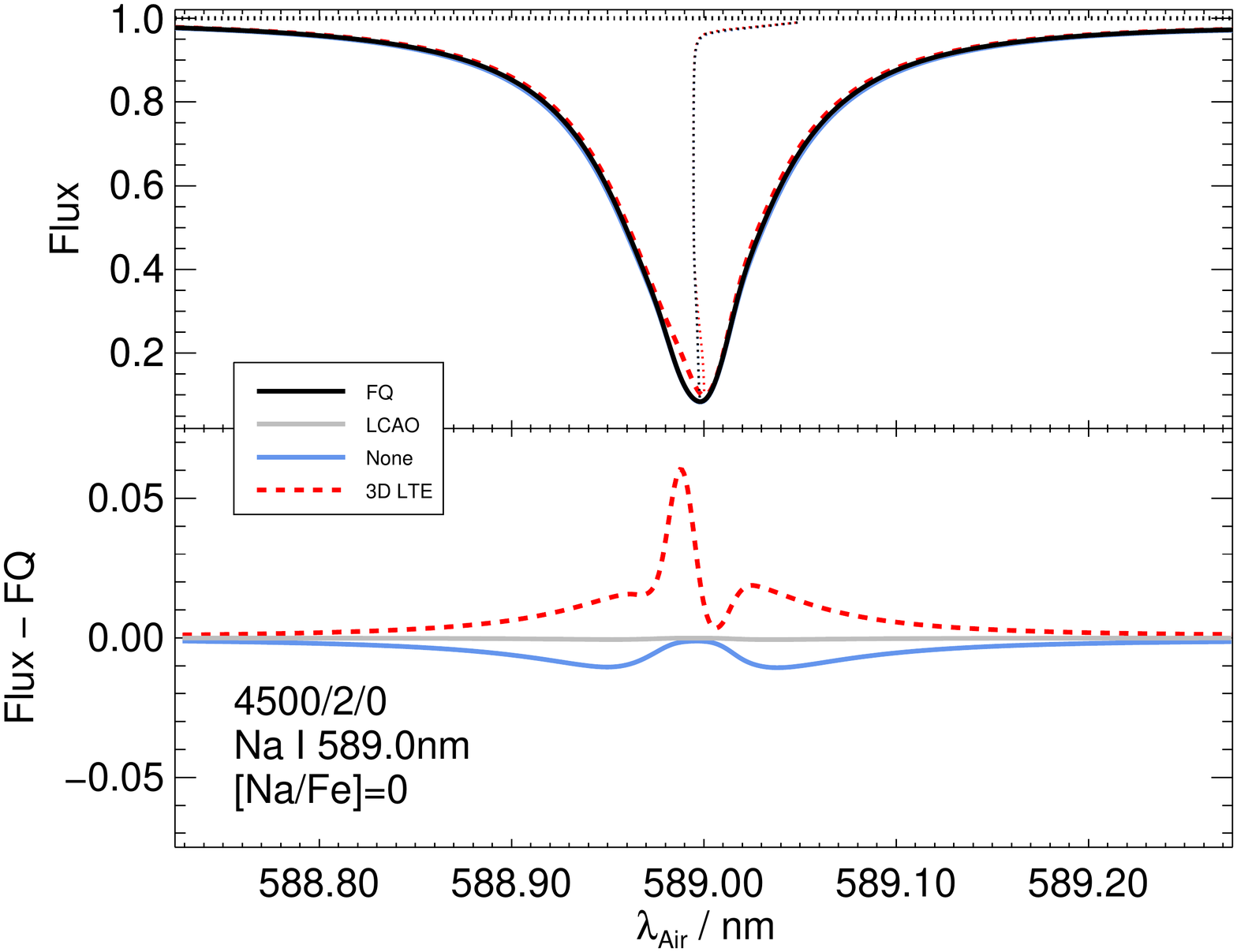}
    \caption{
    Example synthetic 3D non-LTE profiles and bisectors (upper subpanels)
    for the \liline{} (top) and \naline{} (bottom) resonance lines,
    and their residuals with respect to the FQ model (lower subpanels),
    based on different MN descriptions (Table~\ref{tab:modelatoms}).
    Profiles for the models not plotted here
    overlap with FQ and LCAO, and FQ and LCAO are indistinguishable in the line profiles (upper subpanels).
    The 3D \texttt{STAGGER} atmospheres are labelled by 
    \{$\teff/\lgg/\feh$\}. Rotational and instrumental
    broadening have been neglected.}
\label{fig:prof}
\end{figure}

\begin{figure*}[t]
    \includegraphics[scale=0.33]{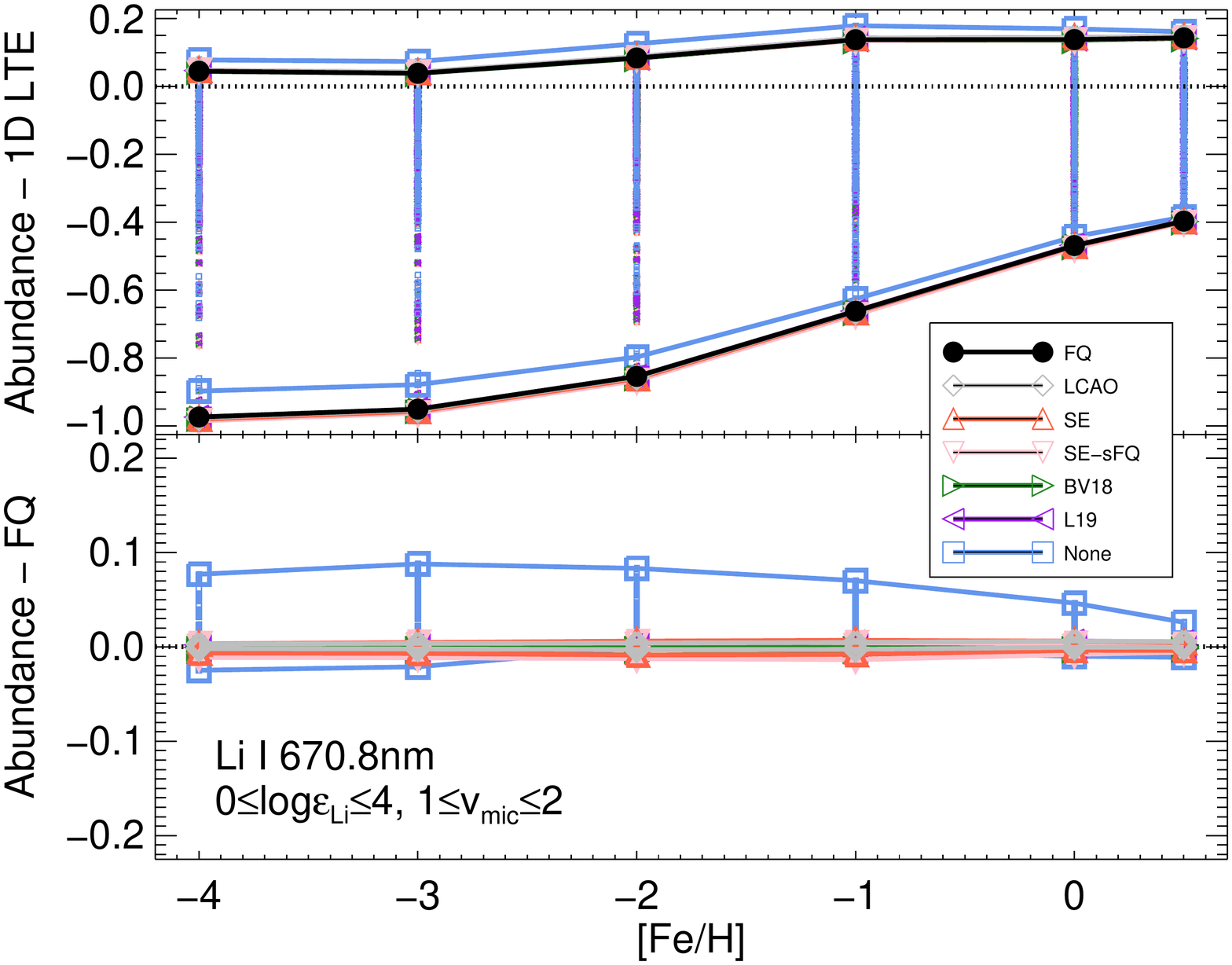}\includegraphics[scale=0.33]{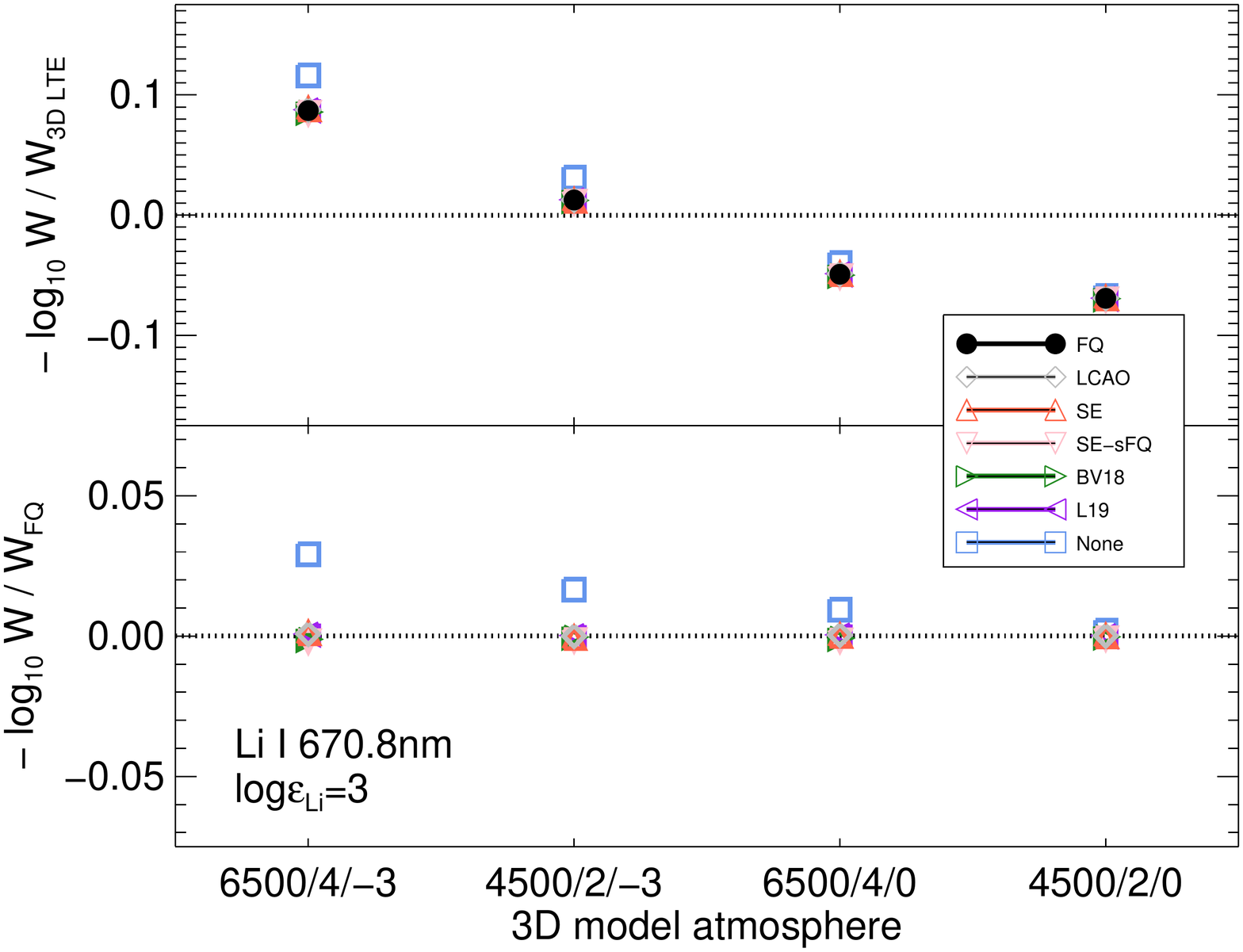}
    \includegraphics[scale=0.33]{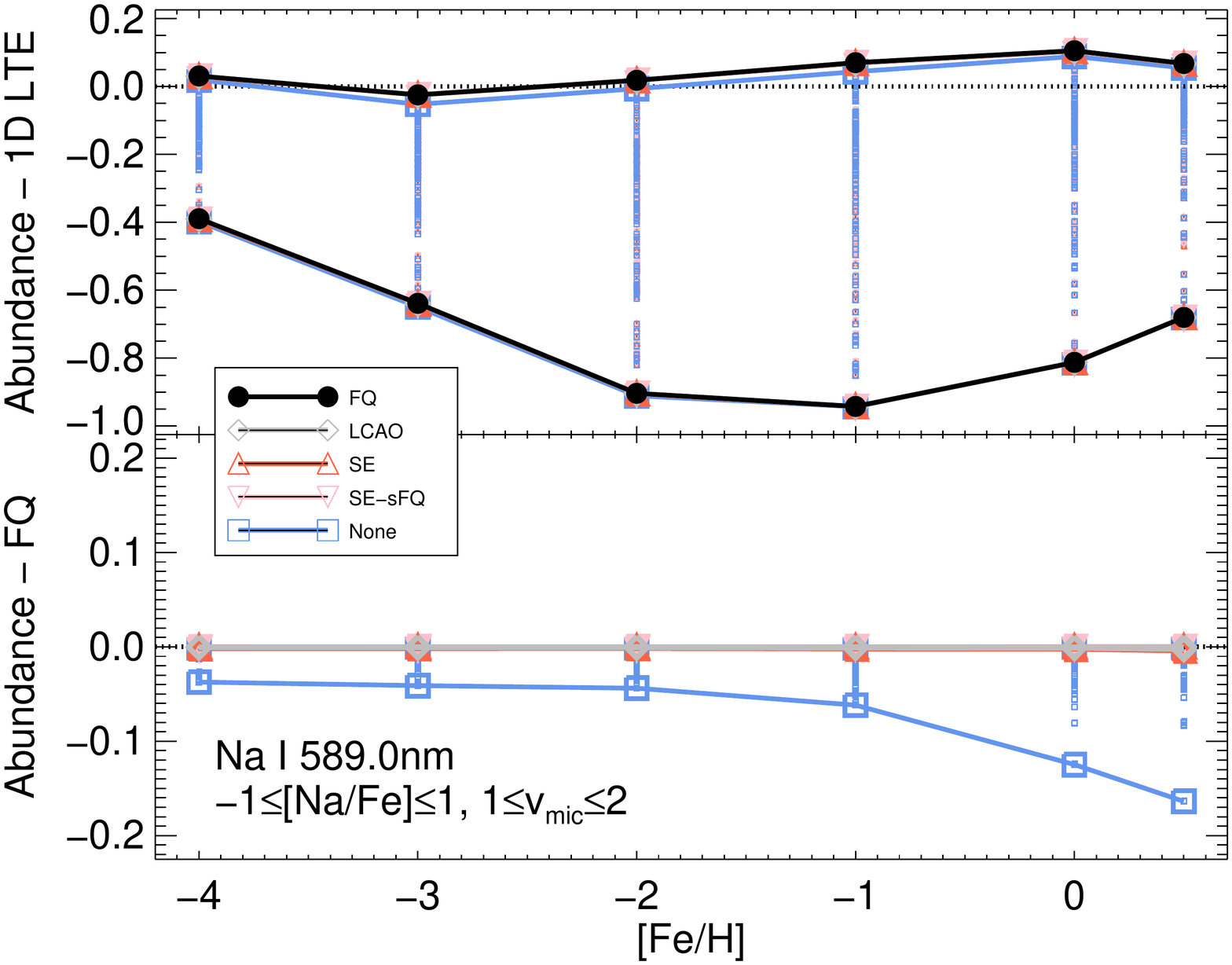}\includegraphics[scale=0.33]{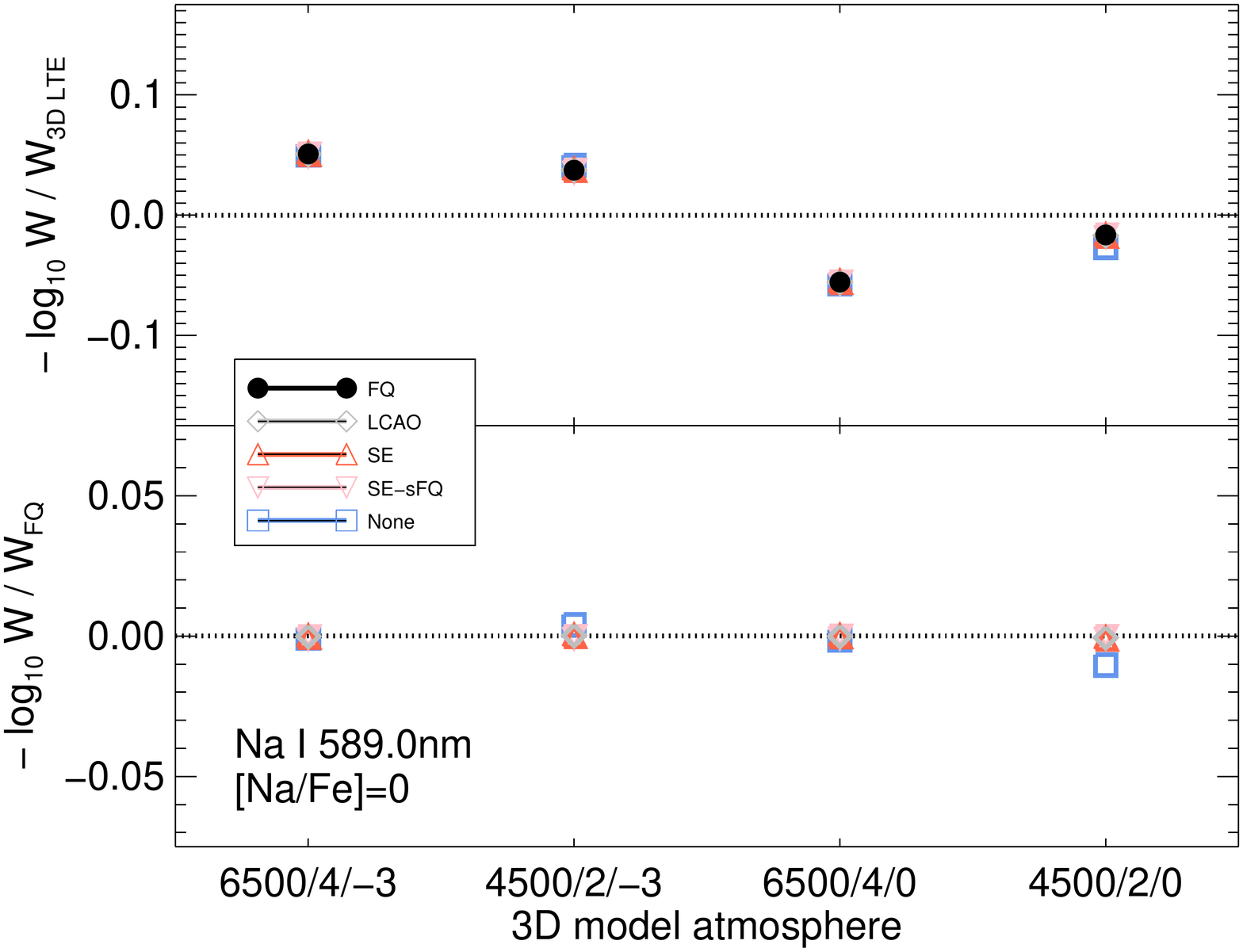}
    \caption{
    Impact of different MN descriptions (Table~\ref{tab:modelatoms}) 
    on strengths of \liline{} (top row) and \naline{} 
    (bottom row) resonance lines.
    Left: 1D non-LTE abundance differences compared to 1D LTE 
    (upper subpanels) and compared to the FQ model (lower subpanels),
    calculated for different
    1D \texttt{MARCS} atmospheres
    using different abundances and microturbulences;
    solid lines delineate the smallest and largest differences.
    Right: logarithmic 3D non-LTE
    equivalent width ratios compared to 3D LTE
    (upper subpanels) and compared to the FQ model (lower subpanels),
    a proxy for the abundance differences;
    the 3D \texttt{STAGGER} atmospheres are labelled by 
    \{$\teff/\lgg/\feh$\}.}
\label{fig:abcor}
\end{figure*}

Based on the study of fundamental MN data from different theoretical models and experiments carried out in the previous section, we now turn to stellar spectrum modelling and the implications for the choice of MN data to be used.  We do this using 1D and 3D non-LTE radiative transfer calculations,
based on different descriptions of MN motivated by the discussion
above, finally comparing the shapes and strengths of the resulting line profiles as well as
the effects on inferred Li and Na abundances.
The discussion is centred on
the \liline{} and \naline{} resonance lines, these lines
typically showing the most pronounced non-LTE effects;
however, the conclusions that we draw are qualitatively similar
for other lines sometimes used as abundance diagnostics, 
for example the \ion{Li}{1} $610.4\,\mathrm{nm}$ and 
$812.6\,\mathrm{nm}$ lines
and the 
\ion{Na}{1} multiplets at $515\,\mathrm{nm}$,
$568\,\mathrm{nm}$,
$615\,\mathrm{nm}$, and
$818\,\mathrm{nm}$.

\subsection{1D/3D non-LTE spectrum synthesis}

Spectrum synthesis was carried out with
the 3D non-LTE radiative transfer code \texttt{Balder} 
\citep{AmarsiEffectivetemperaturedeterminations2018},
a modified version of the \texttt{Multi3D} code \citep{Leenaarts2009}. 
Calculations were performed across a grid of $103$ 1D hydrostatic \texttt{MARCS} model atmospheres
\citep{Gustafsson2008} that span effective temperatures
$4000\leq\teff / \mathrm{K}\leq7000$ in steps of $1000\,\mathrm{K}$,
surface gravities $1\leq\lggu\leq5$ in steps of $1\,\mathrm{dex}$, and 
metallicities $-4\leq\feh\leq0$ in steps of $1\,\mathrm{dex}$\footnote{The logarithmic abundance of an element A is defined with respect to hydrogen: $\lgeps{A} = \log \frac{N_\mathrm{A}}{N_\mathrm{H}}+12$, where $N_\mathrm{A}$ is the number density of A. The standard notation for the abundance ratio A/B relative to the solar ratio is used $[\mathrm{A}/\mathrm{B}] = (\lgeps{A} - \lgeps{A}^\sun) - (\lgeps{B} - \lgeps{B}^\sun)$, where $\sun$ denotes the solar value, and all logarithms are to base 10.}, as 
well as the extremely metal-rich case $\feh=0.5$.
In all cases, these were performed for $0\leq\lgeps{Li}\leq4$ and
$-1\leq\xfe{Na}\leq1$, in steps of $0.5\,\mathrm{dex}$
and for two microturbulence parameters:
$\vmic=1$ and $2\,\mathrm{km\,s^{-1}}$.

In addition, 3D non-LTE spectrum synthesis calculations
were performed on four 3D hydrodynamic
\texttt{STAGGER} model atmospheres \citep{Magic2013}.
The four models were chosen close to the edges 
of the 1D grid discussed above: namely, a giant star
with $\teff\approx4500\,\mathrm{K}$ and $\lgg=2$,
and a turn-off star with $\teff\approx6500\,\mathrm{K}$
and $\lgg=4$; in both cases,
this was done for both solar-metallicity ($\feh=0$),
and for low metallicity ($\feh=-3$), the
latter being where hydrogen collisions are typically
of greater importance \citep{Barklem2011}.
These 3D non-LTE calculations were only performed
for $\lgeps{Li}=3$ (close to the primordial value),
and for $\xfe{Na}=0$.

\subsection{Model atoms}

\begin{deluxetable*}{lcccccc}
\tablecaption{
    Comparisons of Li$^+$+H$^-$ MN rate coefficients
    at 6000~K from various sources used in different model atoms.   \label{tab:rates_Li}} 
\tablewidth{0pt}
\tablehead{
& \colhead{FQ} & \colhead{LCAO} & \colhead{SE} & 
  \colhead{SE-sFQ} & \colhead{BV18} & \colhead{L19} 
}
\startdata
Total rate coefficient [cm$^3$/s] & 
    1.21$\times 10^{-7}$ & 9.82$\times 10^{-8}$ & 8.66$\times 10^{-8}$  & 
    1.21$\times 10^{-7}$ & 1.38$\times 10^{-7}$ & 1.10$\times 10^{-7}$   \\
Ratio to FQ                &     1.00  &     0.81  &     0.72  &
    1.00 & 1.14 & 0.91 \\
Ratio to L19               &     1.10  &     0.89  &     0.79  & 
    1.10 & 1.26 & 1.00 \\
\hline
Branching fraction $3s$    &     0.66  &     0.71  &     0.83  &
    0.83 & 0.68 & 0.68 \\
Branching fraction $3p$    &     0.31  &     0.21  &     0.10  &
    0.10 & 0.29 & 0.25 \\
Branching fraction $3d$    &     0.03  &     0.08  &     0.06  &
    0.06 & 0.03 & 0.07 \\
\enddata
\end{deluxetable*}

\begin{deluxetable*}{lcccc}
\tablecaption{Comparisons of Na$^+$+H$^-$ MN rate coefficients at 6000~K from
    various sources used in different model atoms. \label{tab:rates_Na}}
\tablewidth{0pt}
\tablehead{
 & \colhead{FQ} & \colhead{LCAO} & \colhead{SE} & \colhead{SE-sFQ}
}
\startdata
Total rate coefficient [cm$^3$/s] & 9.47$\times 10^{-8}$ & 
    7.78$\times 10^{-8}$ & 7.36$\times 10^{-8}$ & 9.47$\times 10^{-8}$ \\
    Ratio to FQ             &      1.00   &   0.82   &   0.78 & 1.00\\
\hline
    Branching fraction $4s$ &      0.86   &   0.85   &   0.91 & 0.91 \\
    Branching fraction $3d$ &      0.11   &   0.12   &   0.08 & 0.08 \\
    Branching fraction $4p$ &      0.03   &   0.03   &   0.01 & 0.01 \\
\enddata
\end{deluxetable*}

The different non-LTE model atoms\footnote{A ``model atom'' refers to a collection of all required fundamental data on the atom of interest needed for non-LTE modelling its spectrum, including energy levels and descriptions of all radiative and collisional processes on the atom.  Overviews of the basic structure and spectral lines of \ion{Li}{1} and \ion{Na}{1} are available at NIST \citep{NIST_5.7} and Grotrian diagrams in \cite{BashkinAtomicenergylevels1975}.} considered here are summarised 
in Table~\ref{tab:modelatoms}. The model atoms are 
all based on those of 
\citet{Wang3DNLTEspectral2020b} and \citet{Lind2011}, for Li and Na respectively.
The only differences between the model atoms
are the rate coefficients for MN involving the 
\{$3s$, $3p$, $3d$\} states of Li, and the \{$4s$, $3d$, $4p$\} states of Na.
For each system, we also tested the importance of these three
transitions, relative to other processes involving 
hydrogen collisions.
In brief, we found that:
\begin{itemize}
\item{Out of all the MN processes included in the models, those involving the
\{$3s$, $3p$, $3d$\} states of Li, and the \{$4s$, $3d$, $4p$\} states of
Na, play the dominating role in setting the statistical equilibria,
across the entire parameter space considered here.}
\item{For Li, out of all processes involving hydrogen collisions 
included in the models, MN 
plays the dominating role in setting the statistical equilibria
across the entire parameter space considered here.}
\item{For Na, out of all processes involving hydrogen collision 
included in the models, MN usually
plays the dominating role in setting the statistical equilibria.
The exception is in cool metal-poor dwarfs,
where excitation by hydrogen collisions also plays a significant
role.}
\end{itemize}

The first three model atoms listed in Table~\ref{tab:modelatoms} are based on
rate coefficients predicted by the theoretical FQ, LCAO and SE models.  
We saw in Sect.~\ref{sec:exp}
that these roughly span the range of uncertainty in the
experimental data, and in light of this it is interesting to consider how this
translates to differences in Li and Na abundances.  Next, the SE-sFQ
model atom adopts the FQ total rate, but with the SE branching fractions.
In Figs~\ref{fig:Li_total} and \ref{fig:Na_total}
we saw that FQ gives the highest total cross section 
while SE gives the lowest;
also, in Figs~\ref{fig:Li_bf} and \ref{fig:Na_bf}
we saw that FQ matches the experimental
branching fractions best, whereas SE is less successful.
Therefore, comparison of results from the SE-sFQ model
with those from the FQ and SE models, may give some hints
regarding the importance of branching ratios into different
final states, versus the importance of the total rate coefficient,
in the context of non-LTE modelling.
Finally, to illustrate their overall importance,
model atoms were constructed wherein MN involving the 
\{$3s$, $3p$, $3d$\} states and the \{$4s$, $3d$, $4p$\} states 
were switched off, for Li and Na respectively (model ``None'').

Two more model atoms shown in Table~\ref{tab:modelatoms}
were constructed specifically for Li.
First, the BV18 model atom is based on the recent calculations by
\cite{BelyaevAtomicDataInelastic2018}.
This set of calculations uses 
the same LiH potentials 
\citep{Croft1999a}
as the FQ model, albeit
adopting the quantum probability current method
instead of full quantum dynamical calculations. 
Finally, the L19 model atom is based on the rate coefficient fits
for MN involving the \{$3s$, $3p$, $3d$\} states provided by
\cite{LaunoyMutualNeutralizationLi2019}.
his data seems to be based on their theoretical calculations which best match the experimental results (ACV5Z+G quantum chemical calculations with Landau-Zener dynamics).

To quantify the relative differences between the models,
Tables~\ref{tab:rates_Li} and 
\ref{tab:rates_Na} compare the total MN rate coefficients and branching
fractions between final states at 6000 K. 
Note, due to high-energy contributions, branching fractions may not correspond exactly with those in the low-energy cross sections.
As expected from the total cross
section comparisons, for Li
the FQ model gives a rate larger than the L19 model, at
6000~K 10\% higher.  
Note that at 6000 K, the most probable reduced collision
energy for Li$^+$+H$^-$ is 0.59 eV/amu, which falls around the bend in the cross
sections.  
In general, the FQ model gives larger total rate coefficients
than the asymptotic LCAO and SE models;
the data from BV18 are larger still.

\subsection{Implementation of rates}
The non-LTE calculations with \texttt{Balder} 
take rate coefficients as inputs,
which result from convolving collisional cross-sections with
the Maxwell-Boltzmann distribution at a given temperature.  
For the FQ cases the rate coefficients were taken directly
from the articles referenced in Table~\ref{tab:modelatoms}, and are based on MN cross sections covering an energy range ($E_\mathrm{CM}$) of roughly $10^{-3}$ to $10$~eV in the case of Li, and $10^{-2}$ to $100$~eV in the case of Na. For LCAO and SE calculations, cross sections are calculated between $10^{-10}$ and $100$~eV.  Outside of these regions cross sections are log-log extrapolated in energy when calculating the rate coefficients.  In the case of FQ data, the reverse rate coefficients for ion pair production (IPP) ($\mathrm{Li/Na}(nl) + \mathrm{H} \rightarrow\mathrm{Li/Na}^+ + \mathrm{H}^- $), are calculated from detailed balance relations, while for LCAO and SE they are calculated directly from cross sections, but confirmed to fulfil detailed balance with the corresponding MN calculation.

For all inelastic collisions,
\texttt{Balder} takes as input the rate coefficients for the process in 
one direction only.  
The code calculates the rate coefficients of the reverse processes on the fly,
via the detailed balance relation.
In this way, spurious non-LTE effects in the deep atmosphere
that may occur from errors due to stipulating the rate coefficients
to finite precision, or due to interpolation errors,
are avoided. 
Since MN rate coefficients have a smoother behaviour with temperature than IPP, \texttt{Balder} uses the logarithmic MN rate coefficients on a grid of temperatures, and interpolates at the atmospheric gas temperatures of interest. 
For consistency,
identical temperature grids were adopted for each model:
they span $2000\,\mathrm{K}$ to $8000\,\mathrm{K}$
in steps of $2000\,\mathrm{K}$. Using a set of LCAO data calculated on a finer grid,
the interpolation error at $5000\,\mathrm{K}$
when interpolating in steps of $2000\,\mathrm{K}$
was found to be smaller than $1\%$
(i.e.~smaller than the nominal accuracy in the rate coefficients
themselves).

\subsection{Results}

The main result of this section is that
the different MN descriptions result in nearly identical 
line strengths and derived abundances for both Li and Na.
This can be seen in the 3D non-LTE profiles of the 
\liline{} and \naline{} lines in Fig.~\ref{fig:prof}.
The residual between the FQ and LCAO models 
is less than $0.0005\,\mathrm{dex}$ for the 3D hydrodynamic models
considered here; the agreement between FQ and the other descriptions
are similarly too small to be resolved in Fig.~\ref{fig:prof}.

This result is clearer in
Fig.~\ref{fig:abcor}, in which we plot abundance differences
calculated on the 1D \texttt{MARCS} and 
3D \texttt{STAGGER} model atmospheres.
These abundance differences
are based on equivalent widths of the spectral lines.
For example, for a given 1D LTE abundance,
the 1D non-LTE versus 1D LTE abundance difference
is defined as $\log\epsilon^{\text{1D NLTE}}-\log\epsilon^{\text{1D LTE}}$
such that the 1D non-LTE equivalent width
is equal to that in 1D LTE (see for example 
Sect.~2.6 of \citealt{amarsi_non-lte_2016}).
In 1D, the abundance differences of the LCAO model
with respect to the FQ model
are at most $0.006\,\mathrm{dex}$ for the
\liline{} line, and $0.003\,\mathrm{dex}$ for the \naline{} line.

The second important result 
concerns the overall importance of MN on non-LTE modelling
in cool stars.
The difference between the results for the FQ model,
and the model without MN (``None''),
shows that neglecting MN does in fact have a significant impact
on spectral line profiles (Fig.~\ref{fig:prof})
and on inferred Li and Na abundances (Fig.~\ref{fig:abcor}).
The example 3D non-LTE profiles in Fig.~\ref{fig:prof}
illustrate a noticeable difference between the resulting profiles,
for particular choices of stellar parameters.
Considering the entire parameter space,
the lower subpanels of the left two plots of Fig.~\ref{fig:abcor} 
illustrate that 
neglecting hydrogen collisions can affect abundances by up to
$0.1\,\mathrm{dex}$ for Li, and $0.2\,\mathrm{dex}$ for Na,
at least for some stars and when using 1D model atmospheres.
In summary, it appears to be important to take MN into account
(even though, as discussed above, the
choice of model for the scattering physics -- FQ, LCAO, SE, etc.-- does
not have a large impact on the inferred abundances, at least for Li and Na).

We briefly note that the result of our test with
the SE-sFQ model, to test the relative importance of
the total rate coefficient to the branching fractions,
was inconclusive. For Li, we found that 
SE-sFQ gives results that are in worse average 
agreement with the FQ model compared to SE.
However, for Na we found the opposite, namely that SE-sFQ 
gives results that are in better average 
agreement with the FQ model compared to SE.
It is unclear why this is, but it is not completely surprising
given that these two alkali species suffer from
different non-LTE effects.
This can be seen in Fig.~\ref{fig:abcor}:
towards lower $\feh$, the abundance corrections for Li become
more severe, whereas for Na they become less severe on the whole.

\section{Conclusions}

Comparisons have been made between experimental and theoretical results for MN
processes in Li$^+$+H$^-$/D$^-$ and Na$^+$+H$^-$/D$^-$ collisions at energies
below 1 eV, total cross sections for Li, and branching fractions for both Li
and Na.  The studies were made with theoretical approaches that have been used
to calculate MN data for astrophysical applications in non-LTE modelling of
stellar spectra, in particular the FQ approach, and the LCAO and SE
asymptotic model approaches.  Generally, the comparisons support the expectation that the FQ calculations are
superior to the model approaches.  Scatter among theoretical calculations and
general agreement with experimental results for the Li$^+$+D$^-$ total cross
section would seem to indicate that total MN rates are not uncertain by more
than 30\%.  Just as importantly, the comparisons indicate the uncertainty does
not vary strongly with collision energy i.e. it is likely to be 
constant in the low energy regime.  The comparisons also indicate that the LCAO
model provides more reliable branching fractions than the SE model.  Again,
this is perhaps in line with expectations, as the semi-empirical couplings are
averaged over different experimental and theoretical results available at the
time \citep{Olson1971} and not just for H$^-$/D$^-$, while the LCAO calculations are specific to the case at hand.

The experiments together with the various sets of theoretical calculations
allow us to place constraints on the accuracy of the rates used in non-LTE
modelling.  This in turn constrains the uncertainty in modelled spectra and derived
abundances for Li and Na arising from this source.  Sets of model atoms for Li
and Na that span the various uncertainties in total cross section and
branching fractions were constructed, and used to compute non-LTE spectra and
abundance corrections with respect to LTE across a large grid of 1D model
atmospheres for late-type stars.  For Li the uncertainties do not exceed
0.006 dex (1.4\%), while for Na the uncertainties do not exceed
0.003 dex (0.7\%). Thus, the most important conclusion of this work is that uncertainties in Li and Na abundances in late-type stars due to uncertainties in MN processes are not larger than these values.

In complex atoms, often of great astrophysical interest such as Fe, presently the
LCAO and SE approaches are the only viable option for estimating
hydrogen collision processes.  FQ approaches are in general too computationally demanding to deal with the many excited states of complex atomic systems required for non-LTE modelling. 
It is unclear if the results found here can be extrapolated to complex atoms, but this
level of agreement for simple atoms gives some confidence in the approach for the MN processes.  
Excitation/deexcitation processes due
to hydrogen were found here to be unimportant for Li, yet they were seen to be important for Na in cool metal-poor dwarfs.  Further,
\cite{AmarsiInelasticcollisions7772018} found that some excitation processes on
O are important to reproduce the centre-to-limb variation of O lines in the
Sun, and that the LCAO method alone may underestimate these 
processes; similar results were later obtained
for C \citep{Amarsi3DnonLTEline2019}.
From a
physics point of view, this would not be surprising, as excitation processes
between near-lying states could occur via mechanisms other than the avoided
ionic crossing mechanism.  From an astrophysics point of view, the fact that even relatively similar atoms like Li and Na have different sensitivities, and different non-LTE effects, means it is difficult to draw general conclusions.

\acknowledgments

We thank Xavier Urbain for providing the Louvain and Newcastle experimental
results in electronic form.  
We thank Ella Wang, Thomas Nordlander, and Karin Lind for providing the base model atoms used in this study.  
Significant parts of this work were performed at
the Swedish National Infrastructure, DESIREE (Swedish Research Council contract
No. 2017-00621).  
This work is a part of the project ``Probing charge- and
mass- transfer reactions on the atomic level'', supported by the Knut and Alice
Wallenberg Foundation (2018.0028).  
The non-LTE computations were performed on resources provided (through project SNIC 2019/3-532) by the Swedish National Infrastructure for Computing (SNIC) at the Multidisciplinary Center for Advanced Computational Science (UPPMAX) and at the High Performance Computing Center North (HPC2N), partially funded by the Swedish Research Council through grant agreement no. 2016-07213.
Furthermore, Paul Barklem, Henrik
Cederquist, Henning Zettergren, and Henning Schmidt thank the Swedish Research
Council for individual project grants (with contract Nos. 2016-03765,
2019-04379, 2016-04181, and 2018-04092).  
Paul Barklem, Anish Amarsi, and  Jon Grumer would like to acknowledge financial support from the project grant ``The New Milky Way'' (2013.0052) from the Knut and Alice Wallenberg Foundation.

\bibliography{MyLibrary_laptop}{}
\bibliographystyle{aasjournal}



\end{document}